\documentclass[sn-mathphys-num,iicol]{sn-jnl}

\usepackage{graphicx}%
\usepackage{multirow}%
\usepackage{amsmath,amssymb,amsfonts}%
\usepackage{amsthm}%
\usepackage{mathrsfs}%
\usepackage[title]{appendix}%
\usepackage{textcomp}%
\usepackage{manyfoot}%
\usepackage{booktabs}%
\usepackage{algorithm}%
\usepackage{algorithmicx}%
\usepackage{algpseudocode}%
\usepackage{listings}%
\usepackage[table]{xcolor}
\usepackage{grffile}

\theoremstyle{thmstyleone}%
\theoremstyle{thmstyletwo}%

\theoremstyle{thmstylethree}%

\usepackage{bm}
\newcommand{\F}{\bm{F}}
\newcommand{\X}{\bm{X}}
\newcommand{\U}{\bm{U}}

\newcommand{\review}[1]{\textcolor{black}{#1}}

\begin{document}

\title[Article Title]{Optimizing Metachronal Paddling with Reinforcement Learning at Low Reynolds Number}


\author{\fnm{Alana A.} \sur{Bailey}}\email{alanabailey29@gmail.com}

\author{\fnm{Robert D.} \sur{Guy}}\email{rdguy@ucdavis.edu}

\affil{\orgdiv{Department of Mathematics}, \orgname{University of California Davis}, \orgaddress{\street{One Shields Ave}, \city{Davis}, \postcode{95616}, \state{CA}, \country{USA}}}




\abstract{Metachronal paddling is a swimming strategy in which an organism oscillates sets of adjacent limbs with a constant phase lag, propagating a metachronal wave through its limbs and propelling it forward. This limb coordination strategy is utilized by swimmers across a wide range of Reynolds numbers, which suggests that this metachronal rhythm was selected for its optimality of swimming performance. In this study, we apply reinforcement learning to a swimmer at zero Reynolds number and investigate whether the learning algorithm selects this metachronal rhythm, or if other coordination patterns emerge. We design the swimmer agent with an elongated body and pairs of straight, inflexible paddles placed along the body for various fixed paddle spacings. Based on paddle spacing, the swimmer agent learns qualitatively different coordination patterns. At tight spacings, a back-to-front metachronal wave-like stroke emerges which resembles the commonly observed biological rhythm, but at wide spacings, different limb coordinations are selected. 
\review{Across all resulting strokes, the fastest stroke is dependent on the number of paddles, however, the most efficient stroke is a back-to-front wave-like stroke regardless of the number of paddles.}}





\maketitle

\section{Introduction}\label{sec1}

At low Reynolds number, many commonly observed macroscopic swimming strategies are not effective due to the lack of inertial forces and highly viscous fluid environment. The time-reversibility of Stokes flow means that symmetric swimming strokes are not viable; the Purcell scallop theorem asserts that any reciprocal motion results in no net displacement of a swimmer \cite{purcell2014life}. Thus, microswimmers have adapted many time-irreversible deformations to swim effectively through viscous fluids, using appendages such as cilia and flagella \cite{elgeti2015physics}.

Ciliated microswimmers, such as \textit{Paramecium} and \textit{Volvox}, coordinate the oscillating motion of their cilia such that successive power strokes of adjacent limbs create a time-asymmetric motion, allowing for a net displacement of the swimmer \cite{byron2021metachronal}. This technique is referred to as metachronal paddling, a propulsion strategy in which a metachronal wave propagates through a swimmer's limbs and drives forward motion of the organism. The direction of the metachronal wave in a one-dimensional array of paddles is either symplectic or antiplectic, meaning that the wave propagation is parallel or antiparallel to the direction of a paddles' power stroke \cite{ghorbani2017symplectic}. Antiplectic metachrony is the more commonly observed form of metachrony, which starts the power stroke sequence with the limbs at the back of the organism and propagates a metachronal wave forward across the array of limbs. Antiplectic metachrony has been shown to generate more net fluid flow and minimize drag \cite{knight1954relations}, however, symplectic metachrony is also observed in nature, for example, in the microswimmer \textit{Opalina} \cite{guo2014cilia}. Furthermore, this swimming strategy is not unique to microswimmers; metachronal paddling is observed in organisms across a wide range of Reynolds numbers, including \textit{Paramecium} \cite{elgeti2015physics} (Re $< 1$), ctenophores \cite{byron2021metachronal} (Re $\sim 10-100$), antarctic krill \cite{murphy2011metachronal} (Re $\sim 1000$), and mantis shrimp \cite{garayev2021metachronal} (Re $\sim 10000$), all of which utilize antiplectic metachrony.

The robustness of the stroke across viscous and inertial flow regimes poses a potential use in designing swimming robots for applications such as drug delivery on microscales \cite{kim2016fabrication, ghanbari2011novel} or underwater target tracking at larger scales \cite{chen2024target}. Within the viscous flow regime, studies have investigated the construction and actuation of artificial microswimmers, taking design inspiration from ciliated and flagellated microorganisms and using magnetism or light for actuation \cite{kim2016fabrication, lim2022fabrication}. Studies have also investigated optimal cilia beating patterns \cite{osterman2011finding} and rigid paddle swimming strategies \cite{takagi2015swimming}, both demonstrating high swimming efficiency with metachronal limb coordination.
In the design of swimming robots, swimming efficiently is critical due to energy constraints \cite{ghanbari2011novel}, so prioritizing efficient limb coordinations is essential. Furthermore, it may not be feasible to implement many sets of appendages to swim, so sufficient work must be done by each paddle. Thus, for this study, we consider paddlers with as few as two sets of limbs and examine the optimal propulsion strategies.

We leverage a reinforcement learning approach to study the optimal limb coordination of a microswimmer with sets of rigid paddles placed along an elongated body. Our rigid paddles are unjointed and inflexible, so propulsion is driven solely by time-asymmetries in limb coordination. The simplicity of our model makes it well-suited for reinforcement learning, as we can check the optimality of the solution selected by the learning algorithm against known propulsion strategies. Reinforcement learning has demonstrated success in various locomotion optimization problems in fluid mechanics \cite{garnier2021review}, including the three-sphere swimmer \review{at low Reynolds number} \cite{tsang2020self}, and the multi-link swimmer \review{at low Reynolds number \cite{qin2023reinforcement} and in potential flow \cite{jiao2021learning}.} The design specifications of these reinforcement learning agents incorporate few degrees of freedom to focus solely on the basic principles of low Reynolds number propulsion. These learning agents are placed in a fluid environment and are then able to discern optimal locomotion patterns through exploration and exploitation learning strategies \cite{sutton2018reinforcement}. Endowing swimming microrobots with the ability to self-learn to swim in challenging environments is a growing area of research in machine learning, as microrobots capable of adapting to their surroundings are largely beneficial in applications \cite{jebellat2024reinforcement}. By approaching metachronal paddling at low Reynolds number in the framework of reinforcement learning, we aim to investigate optimal limb coordination patterns and compare them to the commonly observed swimming strokes found in nature.

\section{Methods}\label{sec2}

\subsection{Model}\label{subsec2.1}
The model swimmer is two-dimensional with an elongated body 10 units long aligned in the horizontal direction with rounded semicircular ends of radius one. Sets of equally spaced paddles of length 3 are placed symmetrically along the top and bottom of the body. Paddles on the top and bottom beat symmetrically so that the body does not rotate and the displacement is only in the horizontal direction. The swimmer moves through a fluid at zero Reynolds number by coordinated motion among its paddles which is explored with reinforcement learning, described in the next section.

\begin{figure}[htb]
    \centering
    \includegraphics[width=0.8\linewidth]{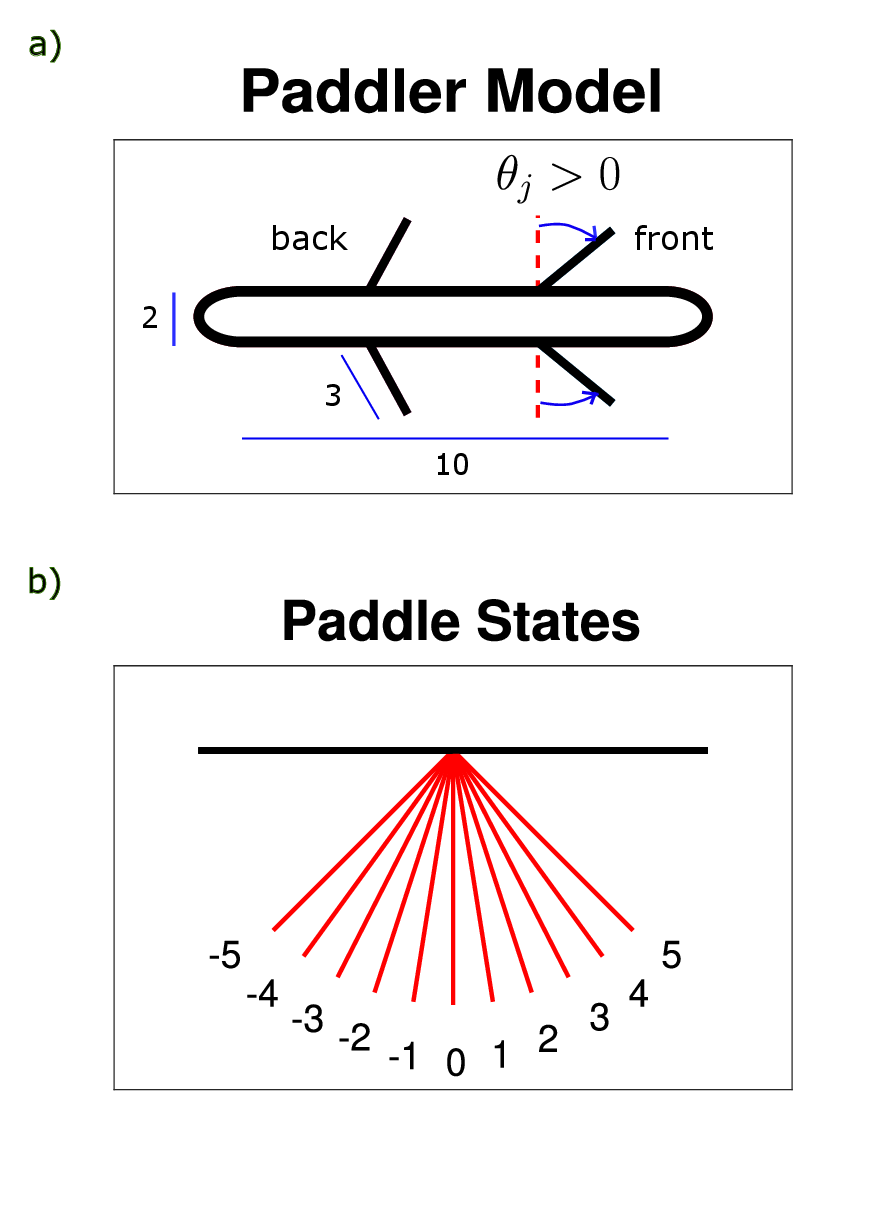}
    \caption{a.) The model paddler has an elongated body of length 10 and pairs of straight, inflexible paddles of length 3 equally spaced along the body. b.) Each paddle can be in one of the above states. States $-5$ and $5$ correspond to a maximum tilt of $\frac{\pi}{4}$ from the center position $0$.}
    \label{fig:paddle_states}
\end{figure}

For a swimmer with $n$ pairs of paddles, the configuration of the paddles is described by the angles $\theta_{j}$ for $j=1\ldots n$, which is related to the angles from the horizontal $\psi_{j}$ of the $j^{th}$ pair of paddles by
\begin{align}
  \psi_{j}^{\text{bottom}} &=  -\frac{\pi}{2} + \theta_{j}, \\
    \psi_{j}^{\text{top}} & =  \phantom{-}\frac{\pi}{2} - \theta_{j}.
\end{align}
A value of $\theta=0$ corresponds to the paddle pair perpendicular to the body (see Figure \ref{fig:paddle_states}).

\subsection{Reinforcement Learning}\label{subsec2.2}
The paddler agent learns to swim on its own by continuously updating its understanding of the fluid environment and the effects of its actions on the environment. To set up the reinforcement learning problem, we must define a state space, action space, and reward function.

The angle of each paddle set, $\theta$, is restricted to the interval $\left [-\pi/4, \pi/4 \right ]$ which is discretized into $11$ equally spaced discrete angles of $\pi/20$. For each paddle, the state is given by the integer $s=-5,-4,\ldots,5$ corresponding to $\theta=s\pi/20$ (see Figure \ref{fig:paddle_states}). For a paddler agent with $n$ limbs on each side, we define our states to be $n$-tuples representing the configurations of the paddles at a given learning step.   


At each learning step, each paddle set can either move one state left, one state right, or not move at all. Hence, the actions are also $n$-tuples representing the paddles' movement information, $a \in [-1,0,1],$ with $-1$ corresponding to moving left, $1$ moving right, and $0$ not moving. The time of each action is one time unit so that the paddles move with constant angular velocity $\dot{\psi}$ of $\pm\pi/20$ or $0$. 

Let $x_{0}$ denote a reference point on the swimmer body. Then $\frac{dx_{0}}{dt} = u_{0}$ is the swimming speed of the paddler. The reward function is defined for each state-action pair as the net displacement of the paddler,
\begin{align}
    r(s,a) = \int_{0}^{1} u_{0} \; dt = x_{0}(1) - x_{0}(0),
    \label{reward_fn:eq}
\end{align}
with forward motion being a positive reward and backward motion being negative.

%
We use tabular Q-learning as our reinforcement learning algorithm due to its simplicity and low computational cost on low-dimensional problems. The estimated quality of each state-action pair, or Q-table, and initial state of the paddler are initialized randomly, then the paddler enters a training loop during which it explores the fluid environment using an $\varepsilon-$greedy strategy, taking mostly random actions initially but prioritizing optimal actions as the training progresses. The estimated quality of each state-action pair is updated at each learning step via the Bellman equation \cite{sutton2018reinforcement}:
\begin{align}
    Q(s_{n},a_{n}) &= (1-\alpha)Q(s_{n},a_{n}) \nonumber\\
    &+ \alpha [r_{n} + \gamma \max_{a_{n+1}} Q(s_{n+1},a_{n+1})].
\end{align}

After the training finishes, we disable any further learning and let the paddler demonstrate the optimal stroke it found. We implement this method on paddlers with 2, 3, and 4 paddles at several different fixed spacings between paddles.

In order to find optimal strokes via Q-learning, we specify the learning parameters in the following way. The learning rate $\alpha$ and the exploration rate $\varepsilon$ are initially set to $1$ and decay geometrically with each learning episode at a rate of $0.99.$ We specify a slow decay rate to allow for a large amount of exploration as the paddler begins interacting with the environment \cite{zhang2024convergence}. In the two-paddle case, the discount factor $\gamma$ is set to $0.99$ and the training loop is run for $50$ episodes with 50,000 learning steps per episode. The 2-paddle swimmer explores a state space with a maximum of 968 states consisting of $11^{2} = 121$ possible paddle configurations and 8 possible actions (there are 3 possible actions per paddle, and we require that at least one paddle moves at every step). Note that with tight paddle spacings, many state-action pairs are not available, so the state space is often smaller than the maximal case. In the three and four-paddle cases, we set $\gamma$ to $0.999$ and run the training loop for $500$ episodes with 500,000 learning steps per episode. Note that each time we add a set of paddles, the number of possible paddle configurations grows by a factor of 11, and the number of possible actions scales as $3^{n}-1,$ where $n$ is the number of paddles. We specify the number of episodes and learning steps to ensure ample time for the paddler to explore the environment and converge to an optimal stroke \cite{sutton2018reinforcement}. These choices of parameters are explored in more detail in section \ref{subsec3.5}.


%
%
\subsection{Fluid Mechanics}\label{subsec2.3}
The fluid motion and resulting translation of the body is determined by solving Stokes equations
\begin{align}
    \mu \boldsymbol{\nabla}^{2}\boldsymbol{u} - \boldsymbol{\nabla}p + \F &= \boldsymbol{0},\\
    \boldsymbol{\nabla} \cdot \boldsymbol{u} &= 0,
\end{align}
where $\F$ is the force on the fluid from the swimmer. We set $\mu = 1,$ noting that $\mu$ scales the forces but does not affect the velocity for problems involving prescribed kinematics. We use the method of regularized Stokeslets \cite{cortez2001method}, a numerical method based on a regularized Green's function, to solve for the coupled fluid/body motion. The body and paddles are discretized into points equally spaced by 0.1. We use the regularization from \cite{cortez2001method} with regularization parameter $\epsilon=0.05$ for our simulations. The velocity $\U(\X_{i})$ at  discrete point $\X_{i}$ is related to the forces on all the other points by 
  \begin{equation}
    \U(\X_{i}) = \sum_{j} \mathcal{S}_{\epsilon}(\X_{i},\X_{j}) \F_{j},
  \end{equation}
where $\mathcal{S}_{\epsilon}(\X_{i},\X_{j})$ is the regularized Stokeslet tensor. We represent this equation as
\begin{equation}
  \mathcal{M}\F = \U,
  \label{FU_MRS:eq}
\end{equation}
where $\U$ and $\F$ represent the collection of velocities and forces, respectively, at all discrete points.

In a given time step, the state-action pair define the body configuration, $\X$, and velocity of deformation, $\dot{\X}$, in a fixed body frame over the unit time interval. Accounting for the motion of the fluid, the overall velocity of the swimmer is the sum of the known prescribed velocity $\dot{\X}=\U_{P}$ and the unknown translational velocity (swimming velocity) $\U_{0}$; i.e.\
\begin{equation}
  \U = \U_{P} + \U_{0}.
  \label{velsum:eq}
\end{equation}
The translational velocity is determined by the constraint that the net force be zero on the swimmer body. Putting together \eqref{FU_MRS:eq} and \eqref{velsum:eq}, with the net force constraint gives the system 
\begin{align}
    \mathcal{M}\F - \U_{0} &= \U_{P}  \\
    \sum_{j}\F_{j}         &= \bm{0},
\end{align}
which is solved at each time instant to determine the forces on each point and the swimming velocity. Given the symmetry of paddling, the swimming velocity is in the $x$-direction and the instantaneous swimming speed is denoted by $dx_{0}/dt$. The reward function, defined by \eqref{reward_fn:eq}, is computed by integrating the instantaneous swimming speed over the time interval required to change paddle configuration using 3-point Gaussian quadrature.


\section{Results}\label{sec3}
We implement the Q-learning algorithm on paddlers with 2, 3 and 4 sets of limbs, varying the spacing between paddles from $0.5$ at the closest to $5$ at the furthest, or otherwise, as far apart as the paddler's body size allows. Depending on paddle spacing, the paddler agent selects different optimal gaits, which we characterize by the timing of the paddles' power and return strokes and the mean position of the paddles. Since the paddler's goal is to move to the right, we define a power stroke as the motion of sweeping through a decreasing sequence of states, or moving a paddle through an arc to the left, which results in net motion to the right. The return stroke is then defined as the motion of sweeping through an increasing sequence of states, or moving a paddle through an arc to the right. Metachronal paddling emerges as a wave of power strokes propagating through the limbs of the paddler.

\begin{figure*}[ht]
    \centering
    \includegraphics[width=1\linewidth]{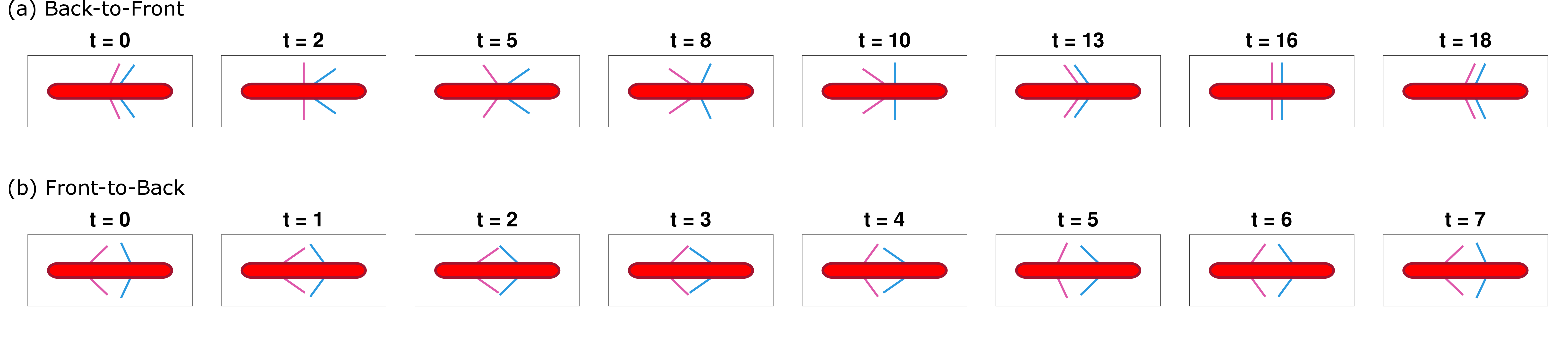}
    \caption{Images of the paddle configuration at different time points for two different strokes for the two-paddle swimmer.  (a) The paddler swims with the back-to-front stroke with paddles spaced $1$ unit apart. (b) The paddler swims with the front-to-back stroke with paddles spaced $4$ units apart.}
    \label{fig:2pad_comic}
\end{figure*}

\subsection{Two Paddles}\label{subsec3.1}
We begin our learning simulations with the simplest paddler capable of swimming at low Reynolds number, a swimmer with two sets of paddles, or a two-paddle swimmer. Depending on the paddle spacing, two distinct strokes emerge: an antiplectic, tilted-out stroke for tight spacings, and a symplectic, tilted-in stroke for wide spacings (see supplemental videos to see the paddler swim with the two strokes.) In particular, the back-to-front stroke is the optimal strategy for paddle spacings $<2,$ and the front-to-back stroke is selected for spacings $\geq 2,$ (see Figure \ref{fig:2pad_comic} for a time sequence of the two strokes).

Qualitatively, the two strokes appear starkly different. To characterize the traits that set these two strokes apart, we compute several stroke metrics, including stroke length, phase lag, and paddle amplitude. Stroke length is defined as the number of moves the paddler makes to complete a cycle, denoted $N.$ For a pair of two adjacent paddle sets, we define the phase lag as the difference in the timing of the power strokes normalized by the stroke length; i.e., for paddle sets $j$ and $j+1$ where set $j$ is left of $j+1$,
\begin{align}
    \Delta \phi = \frac{T_{j} - T_{j+1}}{N}, 
\end{align}
where $T_{j},T_{j+1}$ are the times at which paddle sets $j$ and $j+1$ begin their power strokes, respectively. We report the phase lag in the interval $[-0.5,0.5]$ to capture the shortest time between power strokes, and thus a negative phase lag indicates that the back paddles lead the power stroke sequence. The amplitude of a paddle set is simply the range of motion utilized during the stroke, keeping in mind that the amplitude will always be a multiple of $\pi/20$ due to our discretization.

\begin{figure}[h]
    \centering
    \includegraphics[width=1\linewidth]{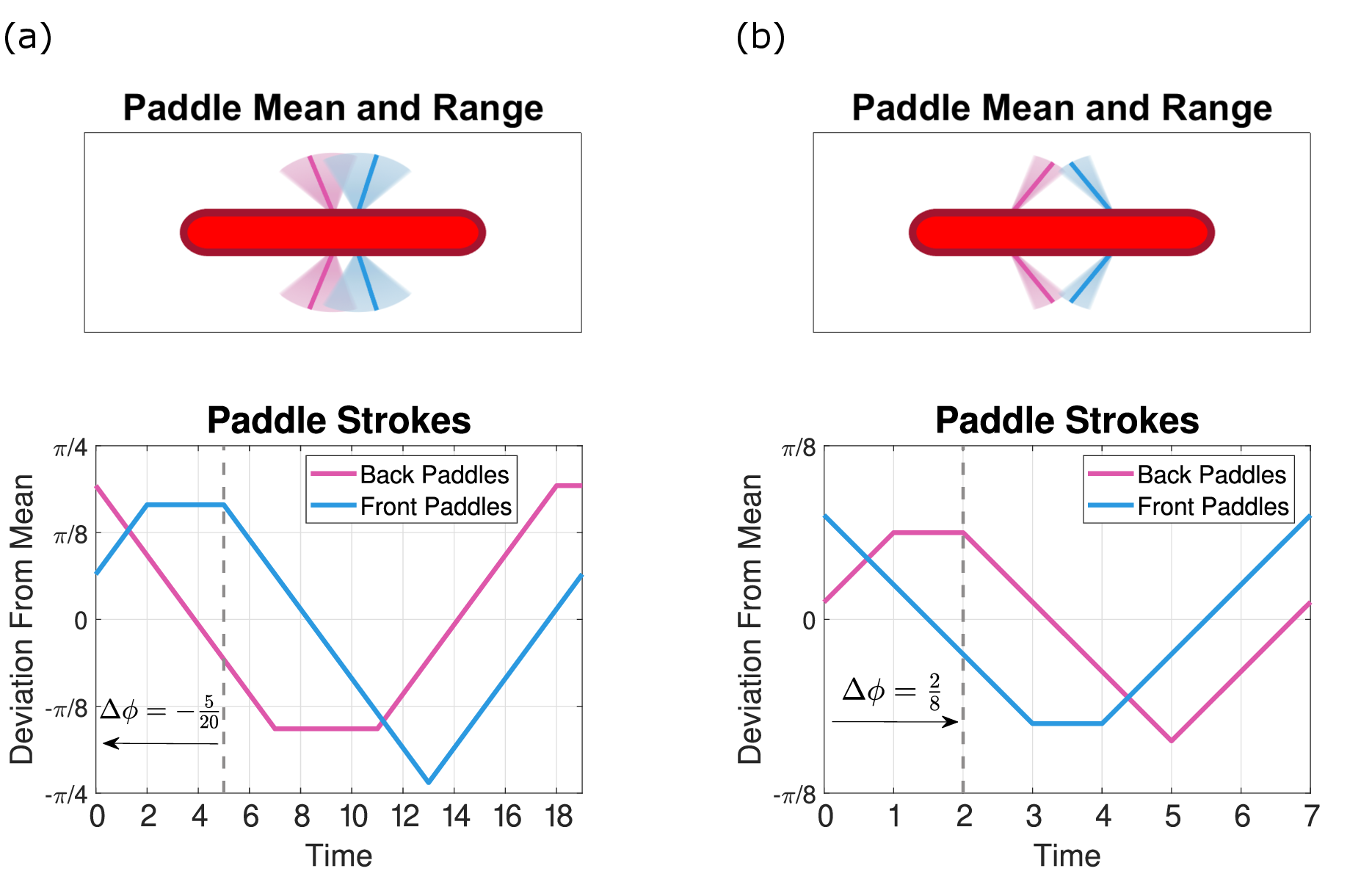}
    \caption{For the two strokes pictured in Figure \ref{fig:2pad_comic}, the top row shows the mean paddle positions (solid line) and the range of paddle motion (shaded region). The bottom row shows the angular displacement from the mean for all paddles over the entire stroke.
    (a) The left column depicts the back-to-front stroke at paddle spacing $1$. We observe a large range of motion of both paddle sets with the means tilted slightly away from each other in the top figure, and the bottom shows $25\%$ phase lag with the back paddles leading (power strokes are shown as negatively sloped lines). (b) The right column depicts the front-to-back stroke at paddle spacing $4$. The top figure shows the inward tilt the paddles maintain throughout the stroke, and the small range of motion used. The bottom figure shows $25\%$ phase lag with the front paddles leading.}
    \label{fig:2pad_mean_ran_tilt}
\end{figure}

\begin{figure*}[ht]
    \centering
    \includegraphics[width=1\linewidth]{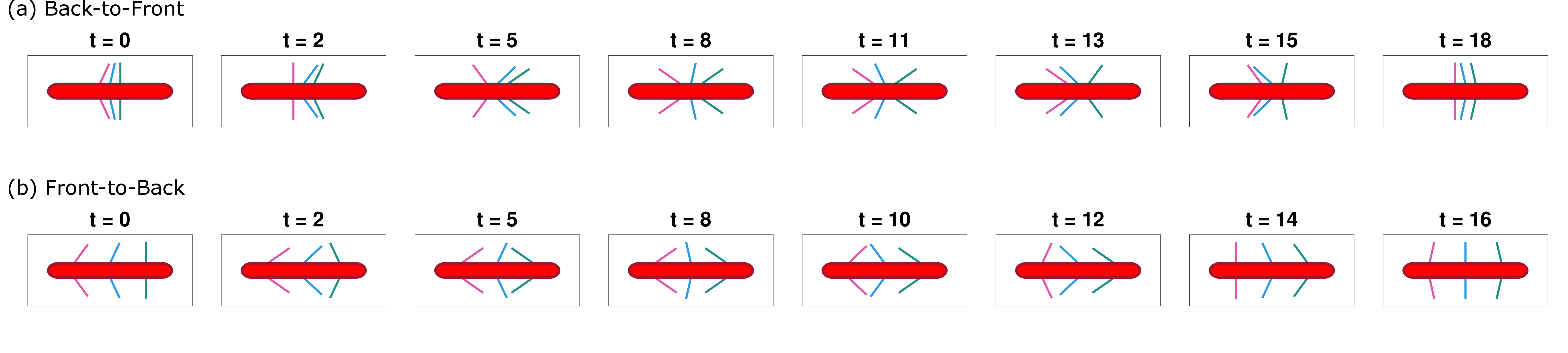}
    \caption{Images of the paddle configuration at different time points for two different strokes for the three-paddle swimmer. (a) The paddler swims with the back-to-front stroke with paddles spaced $1$ unit apart. (b) The paddler swims with the front-to-back stroke with paddles spaced $3.5$ units apart.}
    \label{fig:3pad_comic}
\end{figure*}

Using these metrics to broadly compare the two strokes, the back-to-front stroke demonstrates a longer stroke length and larger paddle amplitudes, while the front-to-back stroke utilizes a much smaller range of motion and a shorter stroke length. Picking one of each stroke type for closer inspection, we investigate a back-to-front stroke at paddle spacing $1$ and a front-to-back stroke at paddle spacing $4$ (see Figure \ref{fig:2pad_mean_ran_tilt} for a visualization of the strokes and Table \ref{tab:2pad} for a side-by-side comparison). For both of these strokes, the phase lag is $25\%,$ but with opposite paddles leading the power strokes. Furthermore, the front-to-back stroke uses only $30\% $ of its paddle range, while the back-to-front stroke uses $70-80 \%.$ In both stroke types, there is some symmetry within the individual paddle strokes; the leading paddles pause at the end of their power stroke, and the trailing paddles pause at the end of their return stroke.

Both strokes in the two-paddle case appear metachronal wave-like, with machine learning selecting either a front-to-back or back-to-front stroke depending on paddle spacing. However, the front-to-back stroke selected at wide spacings is significantly faster. At paddle spacing 4, we see the front-to-back stroke reaching a swimming speed of $0.0605$, while the back-to-front stroke only achieves a speed of $0.0342.$

\begin{table}[h]
    \begin{tabular}{|c|c|c|}
    \hline
    \textbf{Stroke Metric}  & \textbf{Back-to-Front} & \textbf{Front-to-Back} \\ \hline
    Paddle Spacing & 1 & 4 \\ \hline
    Stroke Length & $20$ & $8$ \\ \hline
    Swimming Speed & $0.0342$ & $0.0605$ \\ \hline
    Phase Lag & $-0.25$ & $0.25$ \\ \hline
    Range of Back & $7\pi/20$ & $3\pi/20$ \\ \hline
    Range of Front & $8\pi/20$ & $3\pi/20$ \\ \hline
    \end{tabular}
    \caption{Table comparing the two strokes that emerge from a 2-paddle swimmer at paddle spacings 1 and 4.}
    \label{tab:2pad}
\end{table}

\subsection{Three Paddles}\label{subsec3.2}
With our three-paddle swimmer, we observe similar stroke trends emerging as we vary the paddle spacing. For wider paddle spacings, the paddler performs front-to-back strokes with the outer paddles tilting in towards the middle, and with tighter spacings, a back-to-front stroke with outer paddles tilting outward emerges (see supplemental videos to see the paddler swim with the two strokes.) This back-to-front stroke is optimal for spacings $\leq 2,$ and the front-to-back emerges for spacing between $2$ and $4$ (see Figure \ref{fig:3pad_comic} for a time series of the strokes). However, for spacings larger than $4$, two other non-wave-like strokes emerge that are more challenging to characterize. We examine these other strokes in more detail in Appendix A.

\begin{figure}[h]
    \centering
    \includegraphics[width=1\linewidth]{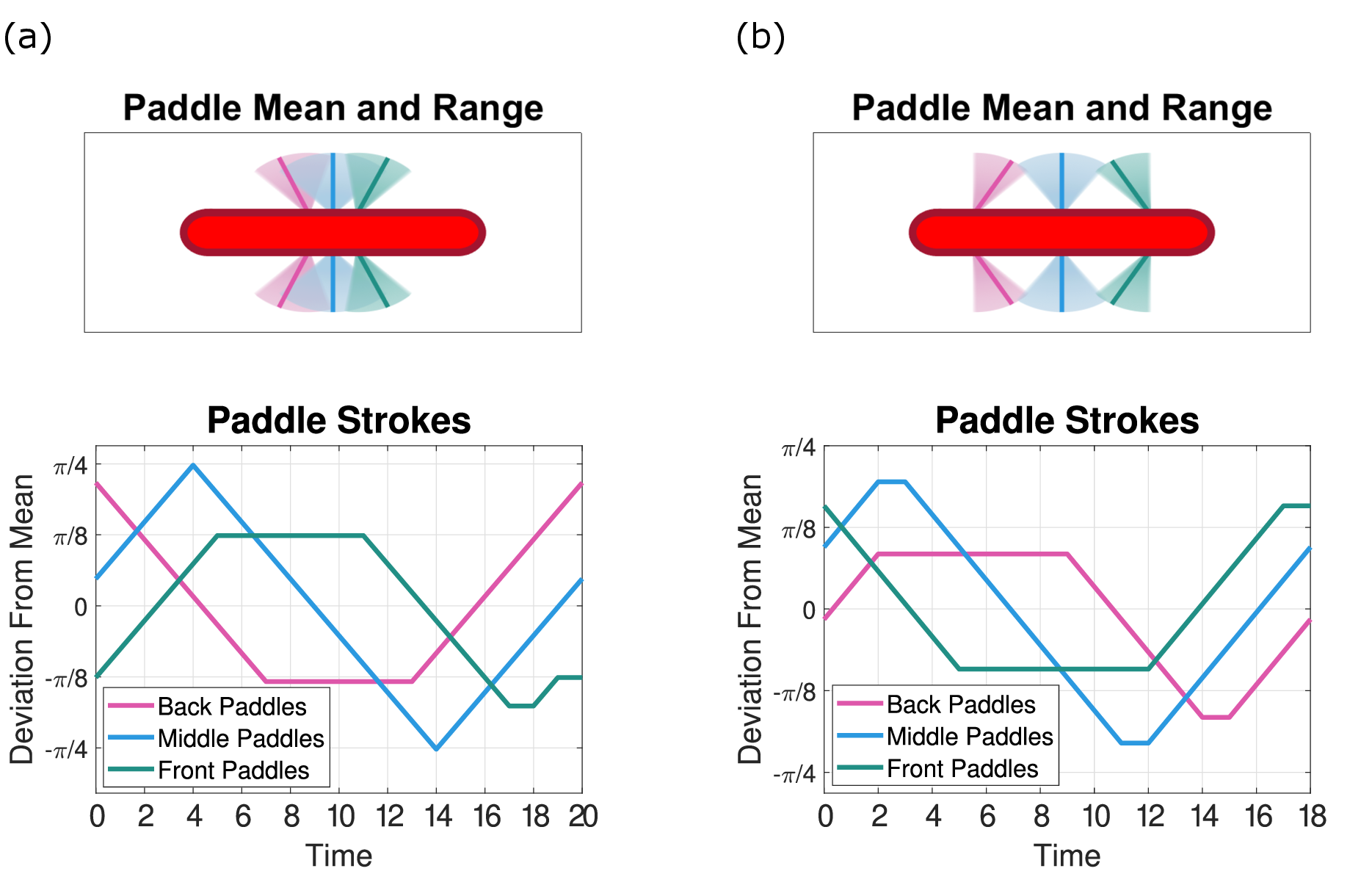}
    \caption{For the two strokes pictured in Figure \ref{fig:3pad_comic}, the top row shows the mean paddle positions (solid line) and the range of paddle motion (shaded region). The bottom row shows the angular displacement from the mean for all paddles over the entire stroke.
    (a) The left column depicts the back-to-front stroke at paddle spacing $1$. We observe a large range of motion of each paddle set with the means tilted slightly away from each other in the top figure, and the bottom shows the back paddles leading and the phase lag between sets of paddles. (b) The right column depicts the front-to-back stroke at paddle spacing $3.5$. The top figure shows the inward tilt of the two outer paddles and the smaller range of motion used compared to the middle paddle. The bottom figure shows the front paddles leading and the phase lags across sets of paddles.}
    \label{fig:3pad_mean_ran_tilt}
\end{figure}

Generically characterizing the two stroke types, the back-to-front tilted-out stroke resembles that of the two-paddle case. The back paddles lead the power stroke followed by the middle and front sets, generally with increasing phase lags across pairs of paddle sets. The front-to-back stroke performs its power strokes in the reverse paddle ordering. For this stroke, the middle paddle sweeps through a large range of motion while the outer paddles tilt inward towards the center and perform smaller movements. 

Examining the paddle spacing $1$ and $3.5$ cases (as shown in Figure \ref{fig:3pad_mean_ran_tilt} and Table \ref{tab:3pad}), we see that for both stroke types, the stroke lengths are comparable, which is a change from the two-paddle case. Next, we compare phase lags across pairs of paddle sets in the order in which they power stroke, finding that the pairs of paddle sets have comparable phase lags and that the phase lag increases from leading to trailing paddles. In terms of paddle symmetry, we again observe in both cases that the leading paddle pauses after the power stroke and the last paddle pauses after the return stroke. As for the amplitude, the back-to-front stroke utilizes a larger range of motion than the front-to-back stroke with each of the paddle sets.

Again, both strokes that emerge are wave-like, with the back-to-front stroke having a more metachronal motion in the symmetry of the individual paddle strokes. Contrary to the two-paddle case, here the back-to-front stroke is the faster stroke, with a swimming speed of 0.0548 compared to 0.0440 with the front-to-back stroke.

\begin{table}[h]
    \begin{tabular}{|c|c|c|}
    \hline
    \textbf{Stroke Metric}  & \textbf{Back-to-Front} & \textbf{Front-to-Back} \\ \hline
    Paddle Spacing & 1 & 3.5 \\ \hline
    Stroke Length & $21$ & $19$ \\ \hline
    Swimmin Speed & $0.0548$ & $0.0440$ \\ \hline
    Phase 1st/2nd & $-0.19$ & $0.16$ \\ \hline
    Phase 2nd/3rd & $-0.33$ & $0.32$ \\ \hline
    Range of Back & $7 \pi/20$ & $5 \pi/20$ \\ \hline
    Range of Mid & $10 \pi/20$ & $8 \pi/20$ \\ \hline
    Range of Front & $6 \pi/20$ & $5 \pi/20$ \\ \hline
    \end{tabular}
    \caption{Table comparing the two strokes that emerge from a 3-paddle paddler at paddle spacings 1 and 3.5. We compare the phase lag of pairs of paddles in the order in which they power stroke, so ``Phase 1st/2nd" means back/middle in the back-to-front case and front/middle in the front-to-back case.}
    \label{tab:3pad}
\end{table}

\subsection{Four Paddles}\label{subsec3.3}
\begin{figure*}
    \centering
    \includegraphics[width=1\linewidth]{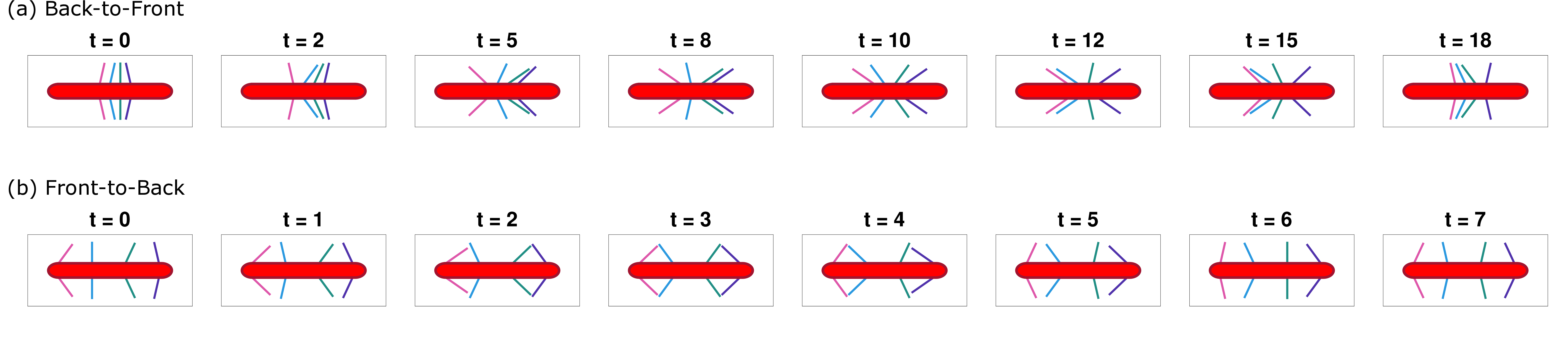}
    \caption{Images of the paddle configuration at different time points for two different strokes for the four-paddle swimmer.(a) The paddler swims with the back-to-front stroke with paddles spaced $1$ unit apart. (b) The paddler swims with the pair-wise front-to-back stroke with paddles spaced $3.25$ units apart.}
    \label{fig:4pad_comic}
\end{figure*}

With four sets of paddles, the paddler performs strokes that resemble the two-paddle case. For tight spacings, we see the familiar back-to-front tilted-out stroke, and with wide spacings, a paired-off front-to-back stroke emerges in which adjacent paddle sets tilt towards one another (see supplemental videos to see the paddler swim with the two strokes.) These paired-off paddle sets perform nearly identical movements with a restricted range of motion resembling the two-paddle case, while the back-to-front stroke continues the trend of large paddle amplitudes (see Figure \ref{fig:4pad_comic} for a time sequence of the strokes).

\begin{figure}[h]
    \centering
    \includegraphics[width=1\linewidth]{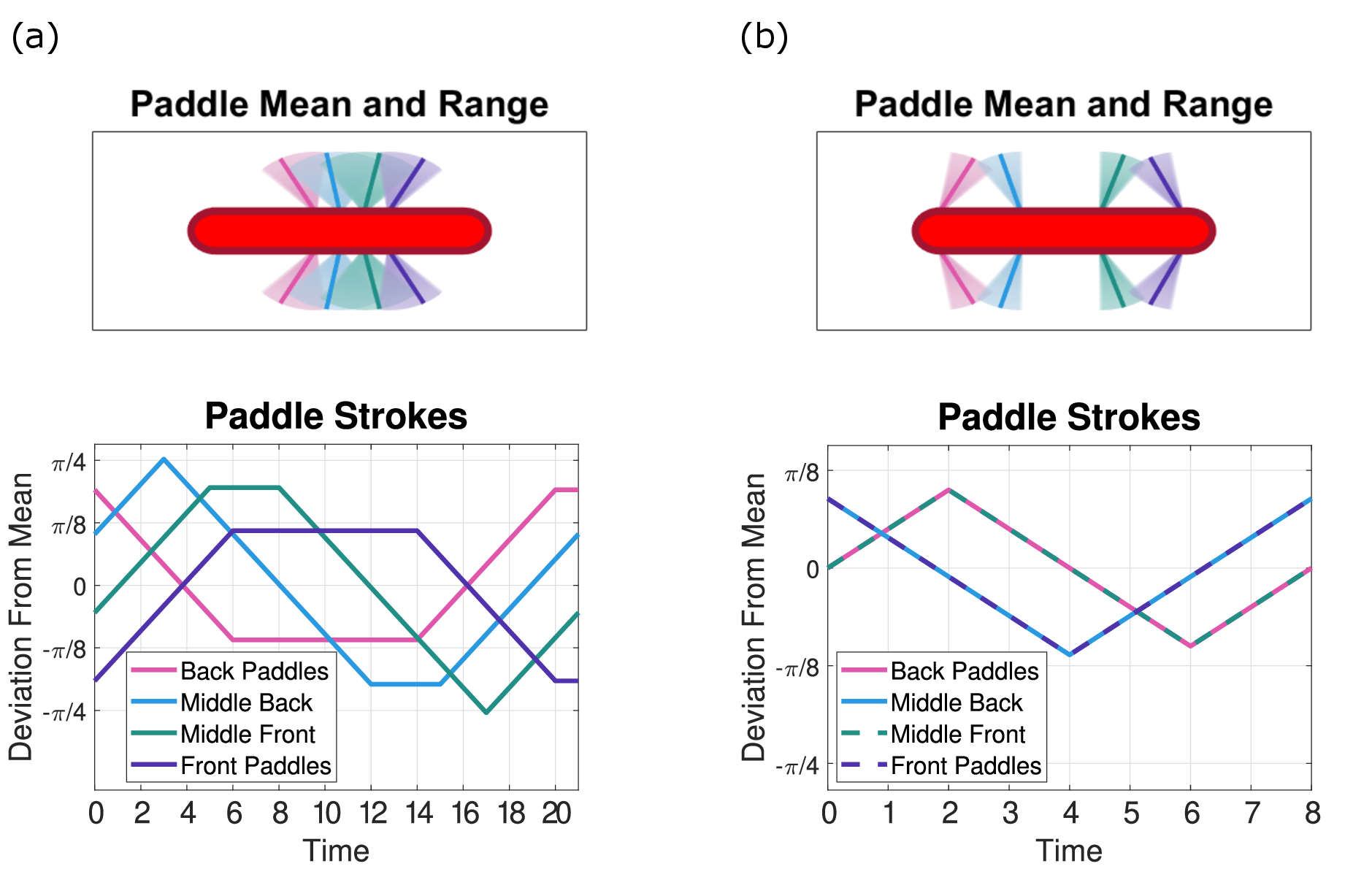}
    \caption{For the two strokes pictured in Figure \ref{fig:4pad_comic}, the top row shows the mean paddle positions (solid line) and the range of paddle motion (shaded region). The bottom row shows the angular displacement from the mean for all paddles over the entire stroke.
    (a) The left column depicts the back-to-front stroke at paddle spacing $1$. We observe a large range of motion of each paddle set with the means tilted slightly away from center in the top figure, and the bottom shows the back paddles leading and the phase lag between sets of paddles. (b) The right column depicts the stroke with paddle's pairing off at paddle spacing $3.25$. The top figure shows the inward tilt the paddle pairs maintain throughout the stroke and the small range of motion used. The bottom figure shows the $22\%$ phase lag between the pairs of tilted paddles, with the front paddles leading.}
    \label{fig:4pad_mean_ran_tilt}
\end{figure}

Comparing the strokes in paddle spacings $1$ and $3.25,$ (visualized in Figure \ref{fig:4pad_mean_ran_tilt} and Table \ref{tab:4pad}), we see that the stroke lengths return to being longer for the back-to-front stroke and shorter for our pair-wise front-to-back stroke. In the back-to-front stroke, the phase lag increases between pairs of paddles starting from the leading set, while the pair-wise front-to-back stroke has a constant phase lag of $0.22$ between the paired-off paddle sets. Similar to before, we observe symmetry in the individual paddle motions. In the back-to-front stroke, the back paddle pauses after the power stroke for the same length of time as the front paddle pauses after the return stroke. The two middle paddles share the same symmetry; the pause length is the same for the back-middle paddle after the power stroke and the front-middle paddle after the return stroke. In the pair-wise front-to-back stroke, we see no pausing, but essentially mirrored versions of the same paddle strokes.

In this case, the back-to-front stroke is metachronal wave-like, however, the front-to-back stroke is overall not wave-like. Moreover, the front-to-back stroke returns to being the faster of the two strokes, with a speed of 0.0818 compared to 0.0680 from the back-to-front stroke.

\begin{table}[h]
    \begin{tabular}{|c|c|c|}
    \hline
    \textbf{Stroke Metric}  & \textbf{Back-to-Front} & \textbf{Front-to-Back} \\ \hline
    Paddle Spacing & 1 & 3.25 \\ \hline
    Stroke Length & $22$ & $9$ \\ \hline
    Swimming Speed & $0.0680$ & $0.0818$ \\ \hline
    Phase 1st/2nd & $-0.14$ & $0.22$ \\ \hline
    Phase 2nd/3rd & $-0.23$ & -0.22 \\ \hline
    Phase 3rd/4th & $-0.27$& $0.22$ \\ \hline
    Range of 1st & $6 \pi/20$ & $4 \pi/20$ \\ \hline
    Range of 2nd & $9 \pi/20$ & $4 \pi/20$ \\ \hline
    Range of 3rd & $9 \pi/20$ & $4 \pi/20$ \\ \hline
    Range of 4th & $6 \pi/20$ & $4 \pi/20$ \\ \hline
    \end{tabular}
    \caption{Table comparing the two strokes that emerge from a 4-paddle paddler at spacings 1 and 3.25. We again compare phase lag across pairs of paddles ordered by their power stroke sequence.}
    \label{tab:4pad}
\end{table}

\subsection{Stroke Performance}\label{subsec3.4}
\begin{figure*}[h]
    \centering
    \includegraphics[width=1\linewidth]{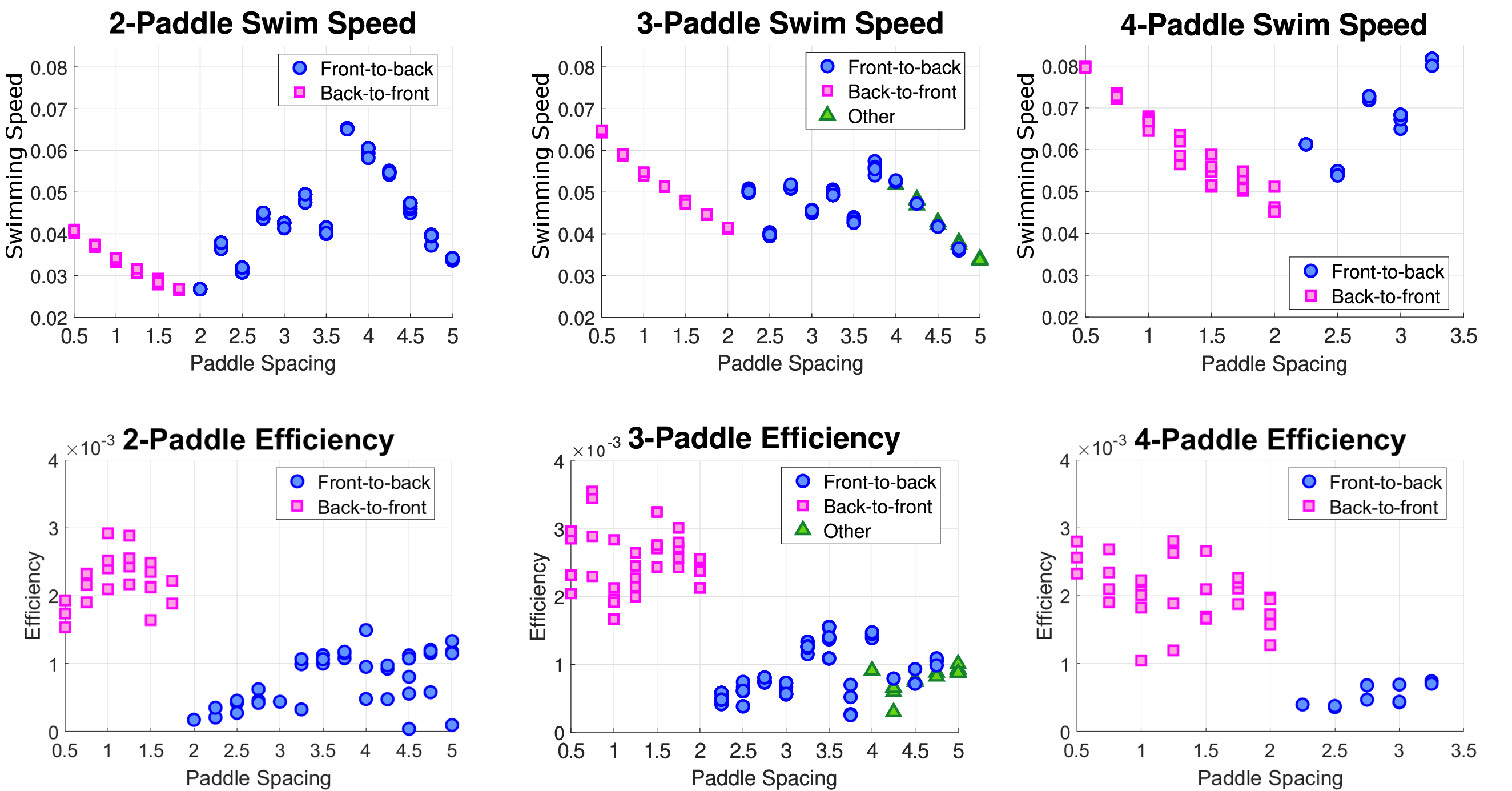}
    \caption{The top row of plots shows swimming speed vs.\ paddle spacing and the bottom row shows plots of stroke efficiency vs. paddle spacing. The blue circles indicate a front-to-back stroke, magenta squares indicate a back-to-front stroke, and green triangles indicate one of the other 3-paddle strokes.}
    \label{fig:6panel_spd_eff}
\end{figure*}

We now evaluate the performance of the machine-learned strokes across the full spectrum of paddle spacings, $0.5$ at the smallest and $5$ at the largest, or otherwise, as far apart as the paddler's body allows. Our performance metrics are the swimming speed and the swimming efficiency, and we compare these metrics across paddle spacings and stroke types (see Figure \ref{fig:6panel_spd_eff}). We run the Q-learning algorithm five times for each paddle spacing with learning parameters that allow for slight stroke variations and plot the resulting swimming speeds and efficiencies from the strokes that emerge.

The swimming speed of a stroke is given by the net displacement of the paddler, or the total reward collected by the paddler agent, normalized by the stroke length,
\begin{align*}
    U = \frac{1}{N} \sum_{n=1}^{N}r(s_{n},a_{n}).
\end{align*}
We quantify the swimming efficiency, $\eta$, by the ratio of the power required to tow the swimmer at the steady swimming speed to the average power used by the paddler over a period:
\begin{align*}
    \eta = \frac{\zeta U^{2}}{P}, \quad P = \frac{1}{N}\int_{0}^{N} \boldsymbol{F}(t) \cdot \boldsymbol{U}(t) \; dt,
\end{align*}
where $\zeta$ is the drag coefficient of the paddler and $\boldsymbol{F}(t)$ and $\boldsymbol{U}(t)$ are the hydrodynamic forces and velocities computed for each move in the sequence. This efficiency metric was originally introduced by Lighthill  \cite{lighthill1975mathematical} and commonly used for zero Reynolds number swimming. More information on the calculation of the drag coefficients is given in Appendix \ref{drag_coeff:app}.

Comparing across the three cases, we see a general trend of the maximum swimming speed increasing as we add more paddles. However, the strokes that achieve the highest speeds vary with the number of paddles. With only two paddles, the front-to-back tilted-in stroke at paddle spacing 3.75 gives the fastest swimming, beating the fastest back-to-front stroke by a factor of $\sim 1.59$. We see the speed increasing with the back-to-front stroke as the space between the paddles decreases, and on the contrary, we see the speed decreasing with the front-to-back stroke as the space between the paddles grows larger than $3.75.$ 

\review{Around paddle spacing 2, we see a transition in stroke type. We investigate both stroke types in detail to determine what is driving this switch, with details provided in \ref{sec:appendixC}. The back-to-front strokes selected at tight spacings are nearly identical, with only slight variation in phase lag and mean angle of the paddles. As we space the paddles farther apart, this stroke becomes less hydrodynamically effective and results in swimming speeds lower than the front-to-back stroke. With the front-to-back stroke, the paddles tilt towards each other to operate near a point of collision; as the space between paddles widens, the mean position of the paddles becomes more tilted inward, and for spacings too narrow, this front-to-back stroke is not viable.}

In the three-paddle case, the fastest stroke is the back-to-front tilted out stroke at the tightest paddle spacing $0.5$. While the front-to-back stroke is dominant for paddle spacing greater than $2,$ it is never faster than the back-to-front stroke with three paddles. Finally, in the case of four paddles, the front-to-back tilted-in stroke returns to being the fastest stroke at paddle spacing $3.25.$ However, this stroke is only faster than the back-to-front stroke at spacing $0.5$ by a factor of $1.02.$ The maximum overall swimming speed is achieved by the 4-paddle swimmer performing the front-to-back stroke at paddle spacing $3.25.$

Now comparing stroke efficiency, we see a common trend across the three cases that the back-to-front stroke is generally more efficient than the front-to-back stroke. This is especially true in the two and four paddle cases, where the maximum front-to-back efficiency is comparable to the minimum back-to-front efficiency. In the three paddle case, there is an efficiency regime where both stokes perform similarly. The maximum overall efficiency is achieved by the 3-paddle swimmer performing the back-to-front stroke at paddle spacing $0.75.$

Recall that the goal of the paddler agent is to optimize swimming speed, so we see little variation in the swimming speeds of the five runs. On the other hand, stroke efficiency is not explicitly being considered by the paddler agent, so there is more variation in the efficiency values.

\begin{figure*}[th]
    \centering
    \includegraphics[width=1\linewidth]{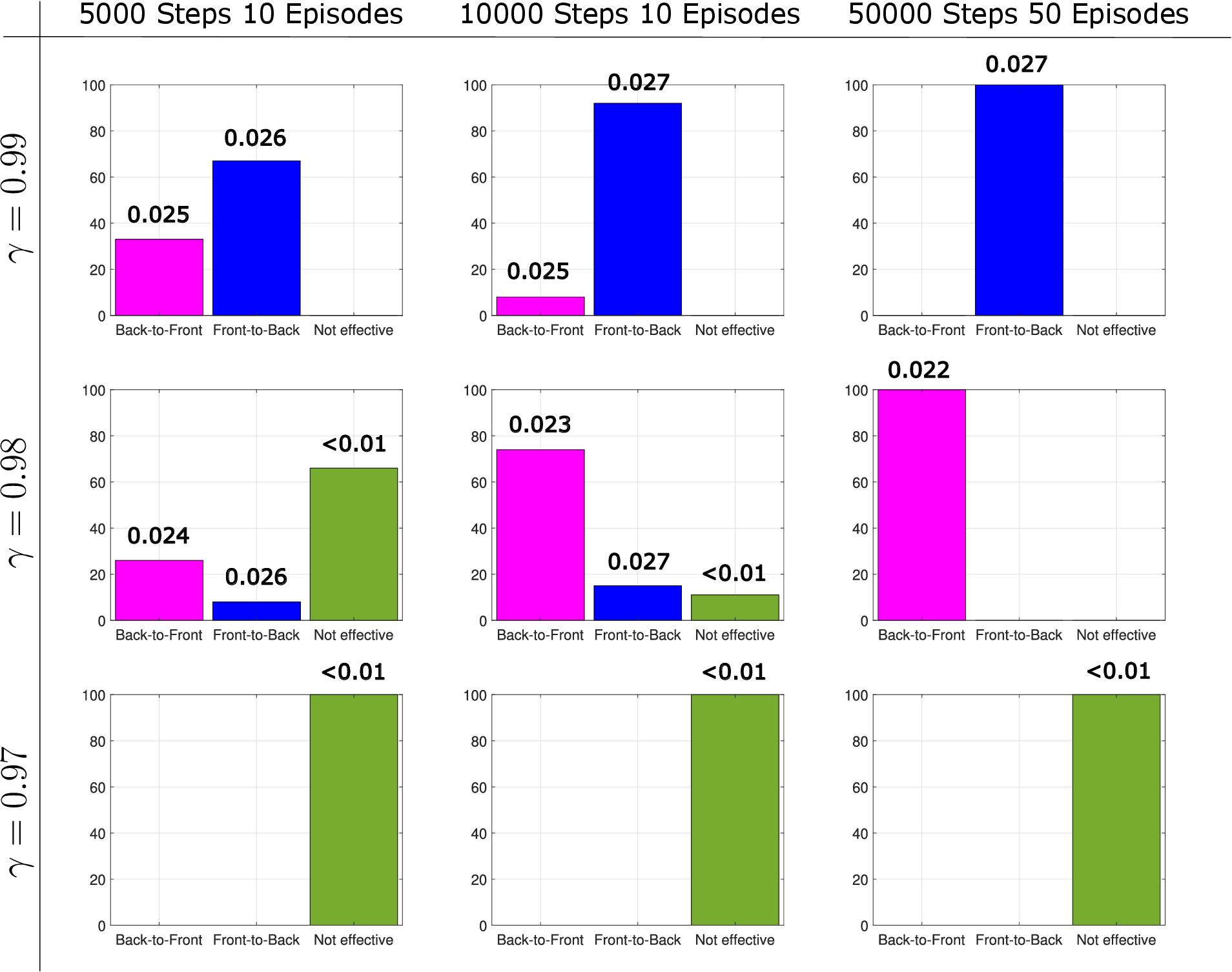}
    \caption{Histograms showing the frequency out of 100 trials of the stroke type found by the 2-paddle swimmer at paddle spacing 2 for various combinations of learning parameters. The numbers above each bar are the mean swimming speed for each stroke.}
    \label{fig:2pad_ml_params}
\end{figure*}

\subsection{Reinforcement Learning Parameters}\label{subsec3.5}
We explore how the learning parameters affect the robustness of the optimal gate produced by the Q-learning algorithm. We first explore the effects of the discount factor and the training loop length on our 2-paddle swimmer with paddle spacing 2, the spacing at which we observe the switch from a back-to-front to a front-to-back optimal stroke. Then, we examine the effects of the exploration rate, $\varepsilon$ and learning rate, $\alpha$, on our 3-paddle swimmer with paddle spacing 2. The 3-paddle swimmer exhibited the most variation in resulting performance and required careful selection of learning parameters.

\begin{figure}[ht]
    \centering
    \includegraphics[width=1\linewidth]{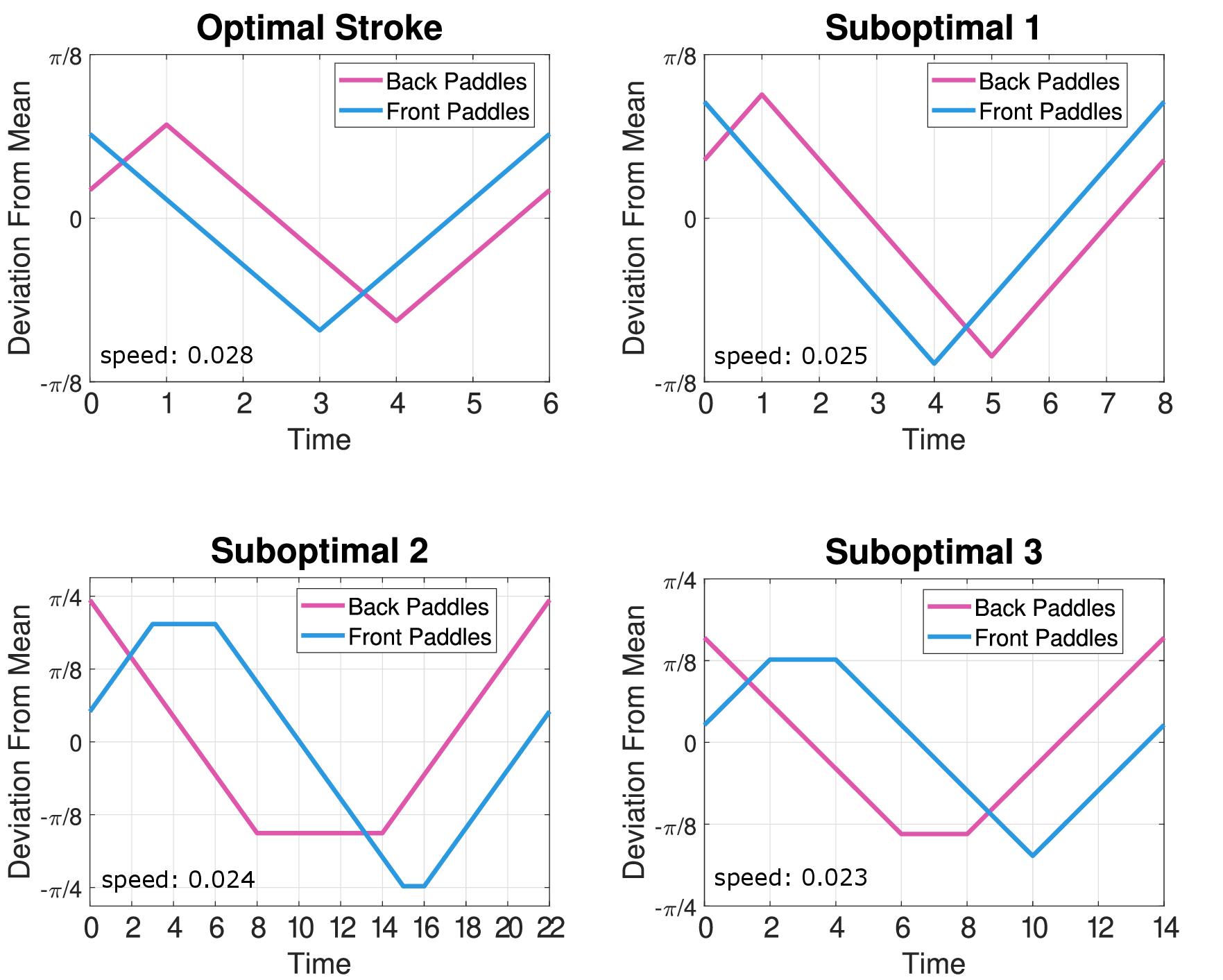}
    \caption{Different swimming gaits produced at paddle spacing 2 for different training parameters. The top row of strokes are front-to-back and are slightly faster than the bottom row of back-to-front strokes.}
    \label{fig:2pad_ml_var}
\end{figure}

\subsubsection{Training Loop Parameters}
First, we examine the effect of varying the length of the training loop. We run the learning algorithm 100 times with $\gamma = 0.99$ and classify the type of stroke for the following combinations of training loop parameters: 10 episodes with 5,000 learning steps, 10 episodes with 10,000 learning steps, and 50 episodes with 50,000 learning steps. 

The first row of Figure \ref{fig:2pad_ml_params} shows how the training loop parameters affect the type of stroke that emerges. With the longest training loop, the paddler converges to a front-to-back stroke every time, but by reducing the number of episodes and learning steps the paddler can learn both front-to-back and back-to-front strokes depending on the randomness of the algorithm (see Figure \ref{fig:2pad_ml_var} for some example strokes). 


With $\gamma$ reduced to $0.98$ (see second row of Figure \ref{fig:2pad_ml_params}), we find that the longest training loop resulted in the paddler converging to a suboptimal back-to-front stroke every time. For shorter training loops we again obtain both stroke types, but we also find that for some initializations, the paddler does not learn an effective propulsion strategy. We classify a propulsion strategy as ineffective if it results in a swimming of less than $0.01.$ For the shortest training loop with $\gamma=0.98$, a majority of the runs resulted in ineffective swimming. We repeat this process once more for $\gamma=0.97$ and observe that for all training loops, the paddler can only learn propulsion strategies that we deem ineffective (see row three of Figure \ref{fig:2pad_ml_params}). 

%

\review{The transition between the two strokes types occurs around paddle spacing two for all numbers of limbs. At this spacing both stroke types emerged depending on the choice of learning parameters.} Other paddle spacings exhibited small variations in the resulting strokes, but did not produce both the front-to-back and back-to-front strokes. For example, at paddle spacing $1,$ there is slight variation in the length of time that the paddles are not moving within a back-to-front stroke, but the paddler does not learn a front-to-back stroke for any known parameter combination. Similar stroke variation trends are observed for the 3 and 4-paddle cases, but due to the larger state space, it becomes computationally infeasible to perform large numbers of trials. \review{Based on this parameter exploration study, we chose to set the discount factor to values close to 1 to ensure sufficient farsightedness of our swimmer, namely $\gamma = 0.99$ for the two-paddle swimmer and $\gamma = 0.999$ for the three and four-paddle swimmers. These values of $\gamma$ are consistent with discount factors used in previous studies on reinforcement learning applied to swimming problems \cite{jiao2021learning, zou2022gait, qin2023reinforcement}.}


\begin{figure}
    \centering
    \includegraphics[width=1\linewidth]{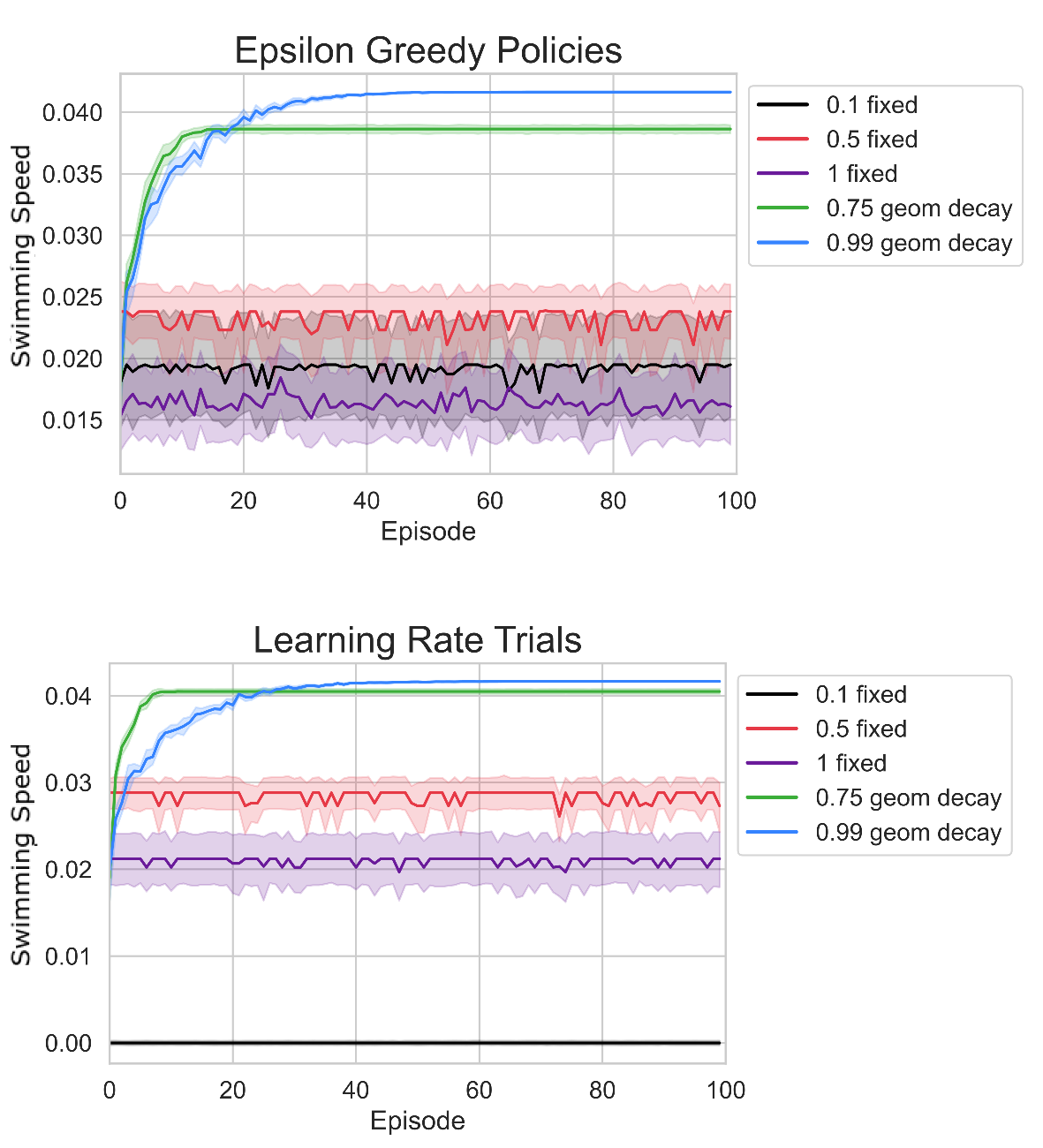}
    \caption{The top figure shows plots of swimming speed vs.\ time for different epsilon-greedy policies with error bars depicting a $95 \%$ confidence interval. The bottom figure shows plots of swimming speed vs. time for different learning rates with error bars depicting a $95 \%$ confidence interval.}
    \label{fig:3pad_decay}
\end{figure}

\subsubsection{Exploration and Learning Rate}
Next, we experiment with our epsilon-greedy policy and learning rate with our 3-paddle swimmer, the optimization landscape that we found most challenging in our simulations. Both the exploration rate and the learning rate decay geometrically in time. It is known that the learning rate $\alpha$ should decay in time for convergence \cite{sutton2018reinforcement}, and similarly, decaying the exploration rate $\varepsilon$ provides better results for stroke optimality \cite{zhang2024convergence}.

 First, we focus on the epsilon-greedy policy and fix the learning rate strategy. The learning rate is initialized at $\alpha = 1$ and decays geometrically with rate $0.99$ per episode (i.e.\ reduced by 1\% per episode). We test the effectiveness of geometrically decaying epsilon-greedy policies against policies with fixed exploration rates. The policies we test include two geometric decay strategies, with rates $0.75$ and $0.99,$ and three fixed epsilon strategies with $\varepsilon = 0.1,$ $0.5$, and $1.$ We run a learning simulation with 300 episodes, 300,000 learning steps each, and discount factor $\gamma = 0.999.$ This simulation is run ten times for each epsilon-greedy policy and the swimming speeds from the resulting strokes are displayed and compared in Figure \ref{fig:3pad_decay}. 
 
 We find that policies with decaying exploration rates result in higher swimming speeds that improve over episodes, and the slower decay rate yields the best performance.  With fixed exploration rates, the swimming speed does not increase with the number of episodes, and the resulting speed is below those obtained with decaying exploration rates. 
 
We next explore strategies to update the learning rate, $\alpha$, over the course of a simulation. We fix the epsilon-greedy policy to geometric decay with rate $0.99,$ which we found to be optimal in the previous study, and we compare different strategies for the learning rate. The results for different learning rates are similar to those for different exploration rates. We find that the slowest decay rate $0.99$ results in the best overall performance, but only marginally over a rate of $0.75$. Both decay strategies result in swimming speeds greater than 0.04, while fixed learning rates show no improvement in swimming speed over time.

\section{Discussion}\label{sec4}
Through a reinforcement learning approach, we identified propulsion strategies for a paddler with 2, 3 and 4 sets of paddles for swimming at low Reynolds number. For microswimmers with arrays of appendages, effective propulsion mechanisms often take the form of metachronal paddling, with the limbs oscillating in a time-delayed sequence to propel the swimmer forward \cite{byron2021metachronal}. With paddles placed tightly-spaced along the body, our 2, 3 and 4-paddle swimmers all self-learned a stroke that resembles an antiplectic wave-like stroke. Of the strokes identified by machine learning, this back-to-front stroke was the most efficient \review{for the 2,3, and 4-paddle swimmers across all spacings}, however, it was not always the fastest.

In the 2 and 4-paddle cases our paddler self-learned a front-to-back stroke in which pairs of paddles tilt in towards each other. Given the inward tilt, this stroke is only practical with sufficient space between paddles, and consequently, the reinforcement learning algorithm converged to this stroke only for paddle spacings 2 or larger. While this is a less efficient stroke, it is often a faster stroke, particularly in the two-paddle case.

\review{A natural question that arises from these results is whether the same antiplectic wave-like stroke would emerge if we instead optimize for swimming efficiency. However, optimizing swimming efficiency is not as simple as replacing the reward metric with the efficiency of each state-action pair \cite{zhang2022learning}. Two approaches to optimizing the swimming efficiency of a stroke include modifying the action space to include longer sequences of actions \cite{zhang2022learning} and modifying the reward function to positively weight swimming speed while negatively weighing work  \cite{lai2025navigation}. These approaches could be adapted to this problem but introduce additional parameters and complexity beyond the scope of the current study.}

Any reciprocal motion results in no net displacement at low Reynolds number, so drag-based propulsion strategies rely on asymmetric motion in the form of time or space asymmetries. Our paddling study concentrates on propulsion mechanisms that are strictly driven by the beating rhythm of the paddles. We designed our swimmer with rigid paddles so that it cannot create asymmetries by changing the shape of its limbs mid-stroke. By allowing only one degree of freedom per paddle in the timing of its movements, we isolate the rhythm-based aspect of low Reynolds number metachronal paddling.

Our reinforcement learning results exhibit the types of asymmetries that can drive forward motion in ciliated microswimmers. The antiplectic wave-like strokes demonstrate variation in the phase lag between paddle sets in both the power and return stroke timing. The strokes with pairs of tilted paddles illustrate another effective type of asymmetry in the mean angle, which has not been observed before, to our knowledge.  

The antiplectic metachronal stroke commonly observed in nature emerged as the most efficient stroke and the fastest for tightly spaced limbs. It is known that in antiplectic metachrony, the swimming efficiency increases with number of limbs \cite{omori2020swimming}. Furthermore, microswimmers in nature deform their cilia to maximize fluid interaction during the power stroke and minimize drag during the return stroke \cite{blake1972model}. Investigating these more complex motions stemming from paddles with bending capabilities, or simply more sets of paddles, will require more sophisticated reinforcement learning methods to handle the rapidly growing computational cost.

\review{With our current discretization, the size of the Q-table scales as $33^{n},$ where $n$ is the number of limbs. Consequently, with even just six sets of limbs, the Q-table has $1.3 \times 10^{9}$ entries. Similarly, if we add hinges to our paddles to capture bending mechanics, we are limited in the same way by the size of the Q-table. A natural extension to the learning algorithm for large state spaces is deep Q-learning, which replaces the Q-table with a neural network \cite{li2017deep}. Deep Q-learning has demonstrated success with complex swimming tasks, such as learning optimal gaits for navigation with a jellyfish-like swimmer \cite{chen2024deep} and learning efficient collective swimming strategies for undulatory swimmers \cite{verma2018efficient}, but is limited to discrete action spaces. Actor-critic methods are more sophisticated reinforcement learning algorithms that allow for continuous action spaces and more precise movements of a learning agent. Actor-critic methods are frequently used in control problems for biological systems \cite{jiao2024deep}, and they have been applied in problems of low Reynolds number locomotion involving navigation \cite{zou2022gait, lai2025navigation, liu2023learning}.}



\begin{appendices}

\section{3-Paddle Non-wave-like Strokes}\label{secA1}

\begin{figure}
    \centering
    \includegraphics[width=0.85\linewidth]{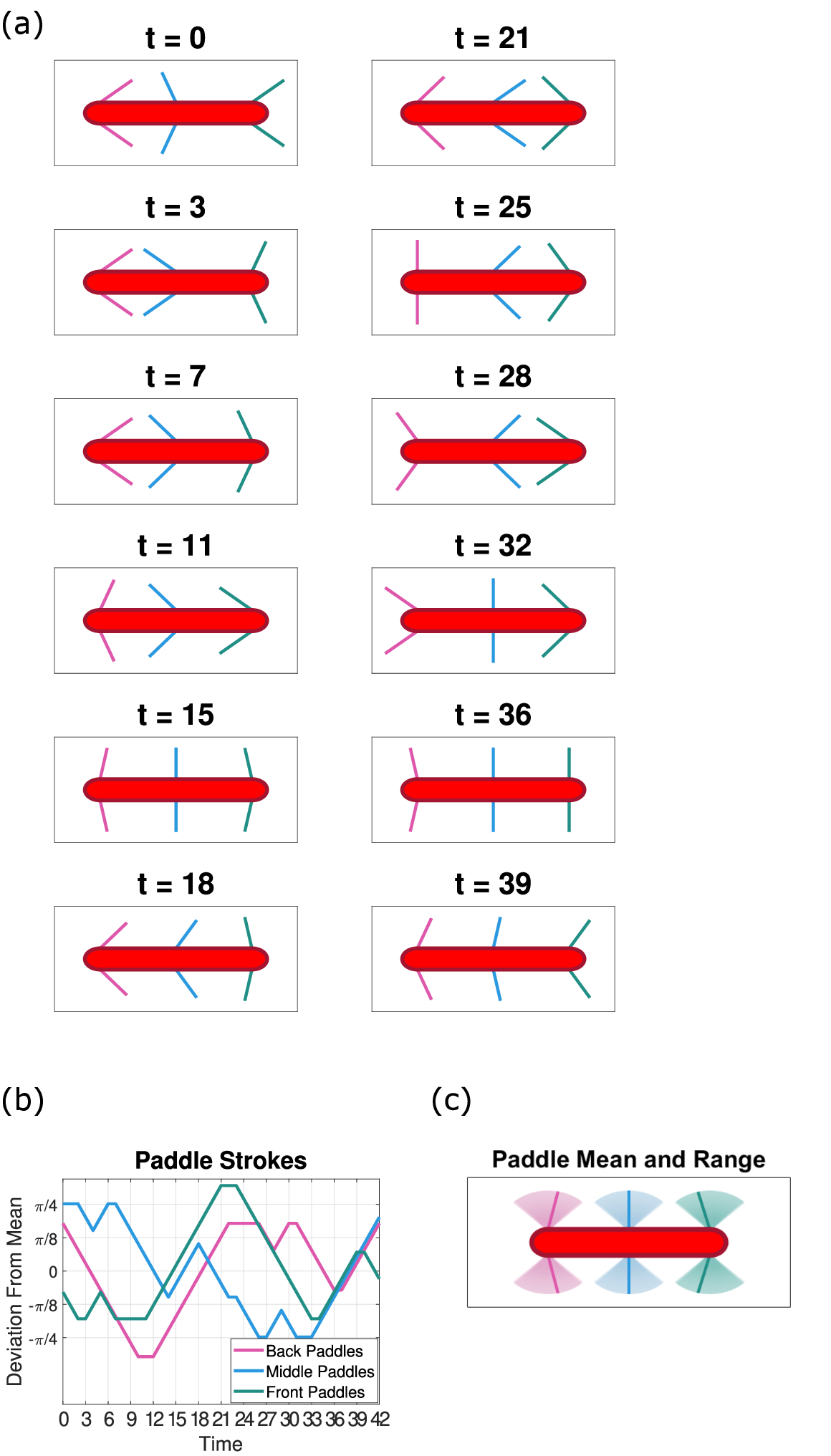}
    \caption{(a) Time sequence of the longer stroke that emerges at the largest spacings. (b) Plot of the paddle angles through time. (c) Mean and range of each paddle.}
    \label{fig:3long}
\end{figure}

\begin{figure}
    \centering
    \includegraphics[width=0.85\linewidth]{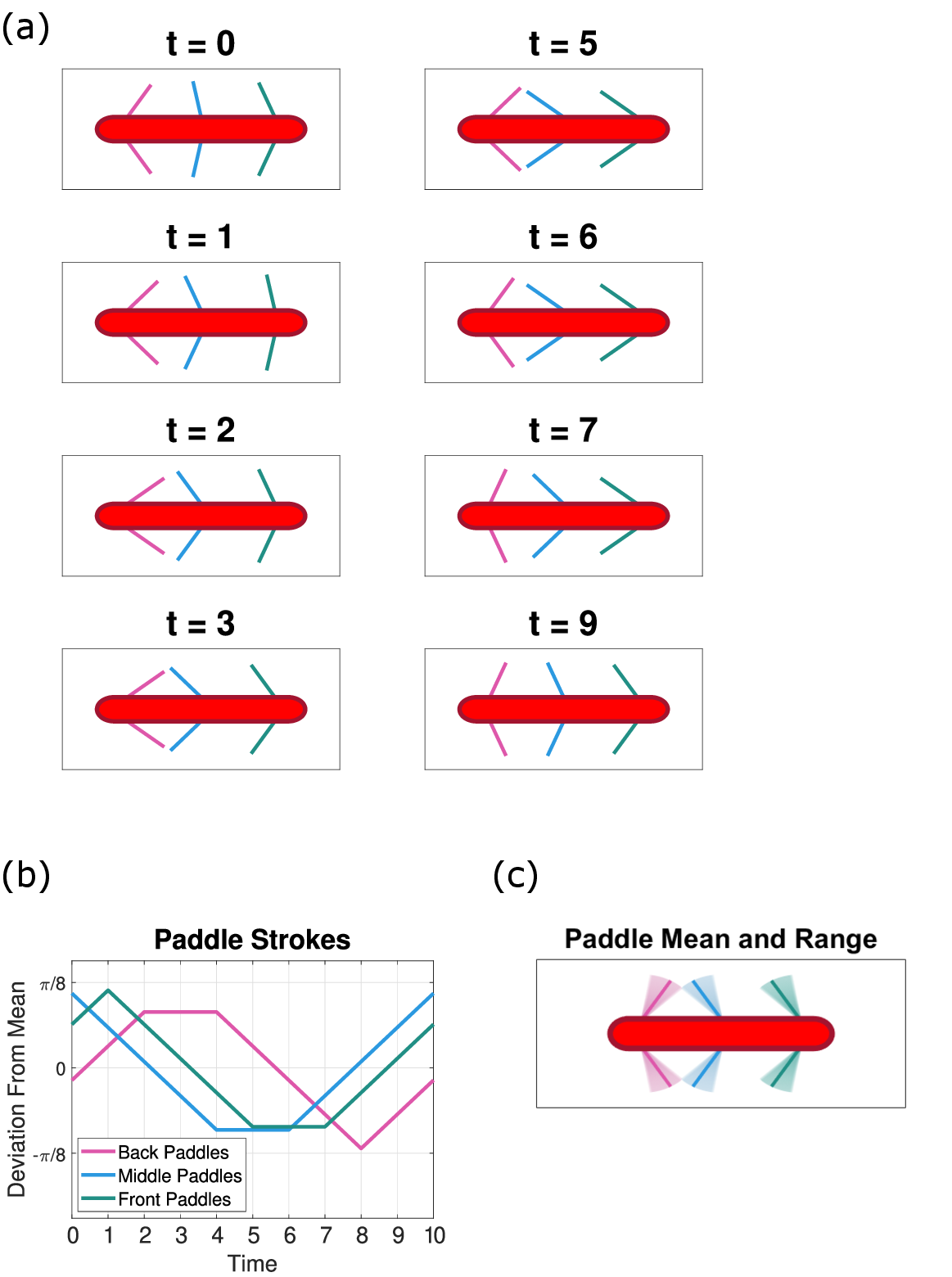}
    \caption{(a) Time sequence of the shorter stroke that emerges at the largest spacings. (b) Plot of the paddle angles through time. (c) Mean and range of each paddle.}
    \label{fig:3short}
\end{figure}

For paddle spacings larger than 4, the three-paddle optimal strokes are more challenging to characterize. One of the strokes that emerges is extremely lengthy compared to previous strokes, taking 43 moves to complete a cycle. The middle paddle alternates between tilting towards each of the outer paddles and timing its movement so that it does two pinching-like motions with each outer paddle (see Figure \ref{fig:3long}). On the other hand, the second stroke we see at wide spacings is very short, with a stroke length of 11; the middle paddle pairs off with one outer paddle and the two remain tilted towards each other for the duration of the stroke, while the other paddle is also tilted inward (see Figure \ref{fig:3short} and see supplemental videos to see the paddler swim with these strokes).

First, considering the stroke of length 43, we see that each paddle utilizes its full range of motion and that the outer paddles have a mean tilt inward. Each paddle undergoes multiple power and return strokes of varying sizes, yet we still see some symmetry with the front and back sets of paddles, as they appear to be mirrored versions of the same stroke.

Now, with the stroke of length 11, we return to much smaller paddle amplitudes, with each paddle set only using an amplitude of $\frac{4\pi}{20}.$ In this case, the middle and front paddles are nearly in sync with their movements, with a phase lag of $0.09$, but with the middle paddles leading. The back paddles then perform the mirrored version of the middle and front paddles' stroke, with a phase lag of $0.36$ between the back and middle paddles.

\section{Drag Coefficients}
\label{drag_coeff:app}
In Table \ref{tab:drag_coeff}, we list the drag coefficients for each limb number and spacing combination. To compute the drag coefficient, we fix the paddles perpendicular to the body and tow the paddler with prescribed unit velocity. We then numerically solve $\mathcal{M}\boldsymbol{F} = \boldsymbol{U}$ for $\boldsymbol{F}$ and sum the resulting forces over the discretization points.
\begin{table}[h]
    \begin{tabular}{|c||c|c|c|}
    \hline
    \textbf{Spacing}  & \textbf{2-Paddle} & \textbf{3-Paddle} & \textbf{4-Paddle} \\ \hline \hline
    $0.50$ & 14.9182 & 14.6285 & 14.3480 \\ \hline
    $0.75$ & 14.7764 & 14.3707 & 13.9740 \\ \hline
    $1.00$ & 14.6398 & 14.1282 & 13.6207 \\ \hline
    $1.25$ & 14.5242 & 13.8970 & 13.2996 \\ \hline
    $1.50$ & 14.4110 & 13.6746 & 12.9886 \\ \hline
    $1.75$ & 14.2993 & 13.4589 & 12.6850 \\ \hline
    $2.00$ & 14.1884 & 13.2488 & 12.3876 \\ \hline
    $2.25$ & 14.0779 & 13.0433 & 12.0966 \\ \hline
    $2.50$ & 13.9673 & 12.8415 & 11.8133 \\ \hline
    $2.75$ & 13.8562 & 12.6432 & 11.5402 \\ \hline
    $3.00$ & 13.7443 & 12.4480 & 11.2798 \\ \hline
    $3.25$ & 13.6500 & 12.2559 & 11.0392 \\ \hline
    $3.50$ & 13.5545 & 12.0670 & \\ \hline
    $3.75$ & 13.4575 & 11.8816 & \\ \hline
    $4.00$ & 13.3587 & 11.7001 & \\ \hline
    $4.25$ & 13.2581 & 11.5231 & \\ \hline
    $4.50$ & 13.1555 & 11.3511 & \\ \hline
    $4.75$ & 13.0510 & 11.1848 & \\ \hline
    $5.00$ & 12.9447 & 11.0245 & \\ \hline
    \end{tabular}
    \caption{Table containing the drag coefficients of the paddlers for each paddle number and spacing combination.}
    \label{tab:drag_coeff}
\end{table}

\section{2-Paddle Stroke Details}\label{sec:appendixC}

\begin{table*}[h]
\centering
    \begin{tabular}{|c|c|c|c|c|c|c|c|}
    \hline
    \textbf{Spacing}  & $\left \langle s_{\text{back}} \right \rangle$ & $\left \langle s_{\text{front}} \right \rangle$ & $A_\text{back}$ & $A_\text{front}$ & \textbf{$\Delta \phi$} & \textbf{Length} & \textbf{Min Dist} \\ \rowcolor[HTML]{f3dcfb} \hline
    0.50 & -2.05 &  1.63 & $7\pi/20$ & $8\pi/20$ & -0.25 & 19           & 0.50 \\ \rowcolor[HTML]{f3dcfb} \hline
    0.75 & -2.05 &  1.63 & $7\pi/20$ & $8\pi/20$ & -0.25 & 19           & 0.75 \\ \rowcolor[HTML]{f3dcfb} \hline
    1.00 & -2.28 &  1.44 & $7\pi/20$ & $8\pi/20$ & -0.26 & 18           & 1.00 \\ \rowcolor[HTML]{f3dcfb} \hline
    1.25 & -2.28 &  1.44 & $7\pi/20$ & $8\pi/20$ & -0.26 & 18           & 1.25 \\ \rowcolor[HTML]{f3dcfb} \hline
    1.50 & -2.28 &  1.44 & $7\pi/20$ & $8\pi/20$ & -0.26 & 18           & 1.50 \\ \rowcolor[HTML]{f3dcfb} \hline
    1.75 & -2.28 &  1.44 & $7\pi/20$ & $8\pi/20$ & -0.26 & 18           & 1.75 \\ \rowcolor[HTML]{abd5ff} \hline
    2.00 &  1.50 & -1.50 & $3\pi/20$ & $3\pi/20$ & 0.29  & \phantom{1}6 & 0.15 \\ \rowcolor[HTML]{abd5ff} \hline
    2.25 &  1.78 & -1.78 & $4\pi/20$ & $4\pi/20$ & 0.10  & 10           & 0.18 \\ \rowcolor[HTML]{abd5ff} \hline
    2.50 &  1.78 & -1.78 & $4\pi/20$ & $4\pi/20$ & 0.10  & 10           & 0.28 \\ \rowcolor[HTML]{abd5ff} \hline
    2.75 &  2.50 & -2.50 & $3\pi/20$ & $3\pi/20$ & 0.14  & \phantom{1}6 & 0.03 \\ \rowcolor[HTML]{abd5ff} \hline
    3.00 &  2.11 & -2.78 & $4\pi/20$ & $4\pi/20$ & 0.20  & \phantom{1}9 & 0.24 \\ \rowcolor[HTML]{abd5ff} \hline
    3.25 &  2.50 & -2.50 & $5\pi/20$ & $5\pi/20$ & 0.18  & 10           & 0.28 \\ \rowcolor[HTML]{abd5ff} \hline
    3.50 &  2.50 & -2.50 & $5\pi/20$ & $5\pi/20$ & 0.18  & 10           & 0.45 \\ \rowcolor[HTML]{abd5ff} \hline
    3.75 &  3.72 & -3.72 & $3\pi/20$ & $3\pi/20$ & 0.38  & \phantom{1}7 & 0.30 \\ \rowcolor[HTML]{abd5ff} \hline
    \end{tabular}
    \caption{Table comparing the strokes that emerge at various spacings for the 2-paddle swimmer. $\left \langle s \right \rangle$ denotes the mean position of the paddles given in terms of the states $s \in [-5,...,5]$. Larger mantitude values of the mean position reflect the larger tilt of the paddle. $A$ denotes the amplitude or range of the paddles, and $\Delta \phi$ denotes the phase lag. We also display the stroke length and the minimum distance between paddles for each stroke. The back-to-front strokes are shown in pink and the front-to-back strokes are shown in blue. We see that the back-to-front strokes are nearly identical, with only two distinct strokes emerging that have little variation. The front-to-back strokes have more variation, but we see a general trend of increasing tilt as we increase the paddle spacing and a consistently small minimum distance between paddles.}
    \label{tab:2pad_details}
\end{table*}

\begin{figure*}
    \centering
    \includegraphics[width=\linewidth]{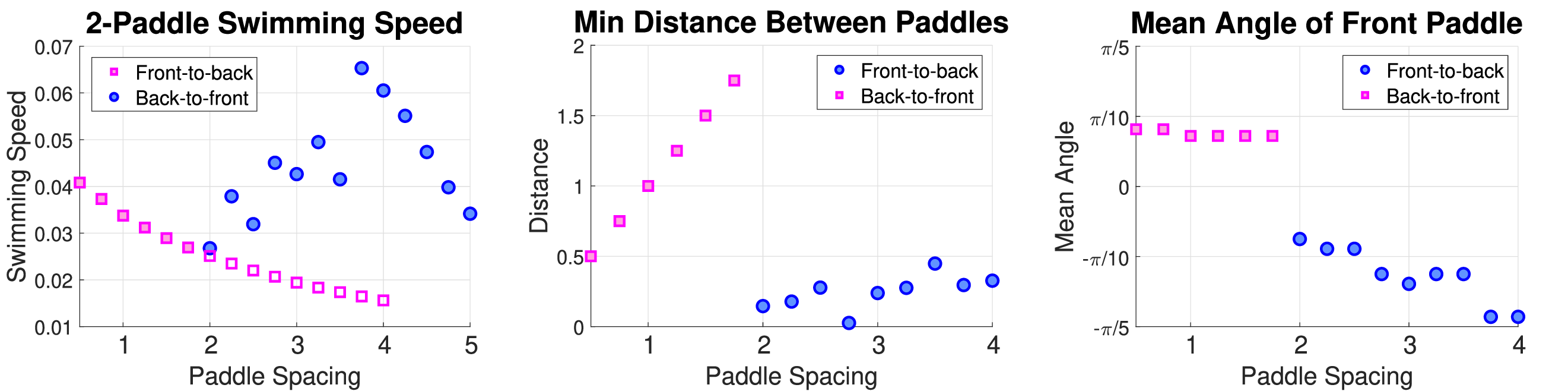}
    \caption{The leftmost plot shows the swimming speed vs.\ paddle spacing for the two-paddle swimmer. Solid markers correspond to the strokes produced by the learning algorithm, and the open dots correspond to the back-to-front stroke extended to wider spacings. The center plot shows the minimum distance between paddles for the selected strokes at each paddle spacing. The minimum distance for the back-to-front stroke corresponds to the paddle spacing at the base, while for the front-to-back stroke, the minimum distance is much smaller than the paddle spacing due to the inward tilt of the two paddles. The rightmost plot shows the mean angle of the front paddle for each paddle spacing. The mean angle is approximately constant for the back-to-front stroke, but the front-to-back stroke shows a steady increase in the inward tilt as the paddle spacing increases.}
    \label{fig:2pad_analysis}
\end{figure*}

\review{In Section \ref{subsec3.4}, we examine the performance of the strokes selected by machine learning using swimming speed and swimming efficiency as metrics. In the 2-paddle data presented in Figure \ref{fig:6panel_spd_eff}, we use a training loop with 10 episodes and 10,000 learning steps, an intentionally short loop to observe the different strokes that could emerge at each spacing. In this section, we use a training loop with 50 episodes and 50,000 learning steps to ensure convergence to an optimal gait, and we more deeply analyze the stroke that emerges as optimal at each paddle spacing. In Table \ref{tab:2pad_details}, we show the mean paddle positions, stroke amplitudes, phase difference between power strokes, stroke length, and the minimum distance between the two paddles over the whole cycle for the 2-paddle swimmer at different spacings. The rows colored pink correspond to the back-to-front strokes, and the rows for the front-to-back strokes are colored blue.}

\review{
The back-to-front strokes are nearly identical for all spacings with only slight variation in phase lag, mean position and stroke length.  As the spacing increases, the swimming speed of this stroke decreases. At spacing 2 and above, the front-to-back stroke was selected by the learning algorithm.  We compute the swimming speeds of the back-to-front strokes at spacings greater than 2, and plot the results in Figure  \ref{fig:2pad_analysis} (left panel). At spacings above 2, the back-to-front stroke resulted in much slower swimming than the front-to-back stroke.}

\review{
The front-to-back stroke changes as the spacing changes. As the spacing increases, the inward tilt of the paddles increases, and so does the swimming speed. The front-to-back stroke always operates near a collision point, as shown by the near-constant minimum distance between the two paddles. The amount of inward tilt is thus limited at tight spacings, which results in slower swimming speeds.}

\review{
We see similar trends with the 3 and 4-paddle swimmers, in that a change in gait occurs around the same paddle spacings. We infer that this gait change occurs for the same reasons as in the 2-paddle case. For wider paddle spacings, a back-to-front stroke is not as effective as a front-to-back stroke, and at tighter paddle spacings, the signature inward tilt of the front-to-back strokes is not strong enough to be effective.}

\paragraph{Supplementary Information}
Animations of the strokes discussed in this manuscript are provided at the following url \url{https://ucdavis.box.com/s/gzsz2zv4424daspsln9x1y3bokpz13mf}.

\paragraph{Author Contribution Statement}
RDG and AAB equally contributed to the design of the study and development of ideas. AAB wrote the computer code, performed simulations, and analyzed data. AAB wrote the manuscript with help from RDG. 

\paragraph{Data Availability}
The data generated in this study are
available from the corresponding authors upon reasonable request.

\paragraph{Open Access}
This article is licensed under a Creative Commons Attribution 4.0 International License, which permits
use, sharing, adaptation, distribution and reproduction in any medium or format, as long as you give appropriate credit to the original author(s) and the source, provide a link to the Creative Commons license, and indicate if changes were made. The images or other third party material in this article are included in the article’s Creative Commons license,
unless indicated otherwise in a credit line to the material. If material is not included in the article’s Creative Commons license and your intended use is not permitted by statutory regulation or exceeds the permitted use, you will need to obtain permission directly from the copyright holder. To view a copy of this license, visit \url{http://creativecommons.org/licenses/by/4.0/}.




\end{appendices}

\bibliography{sn-bibliography}

\end{document}